\documentclass[12pt]{article}
\usepackage{a4wide,epsfig,psfrag,amsmath,amssymb,cite,scalefnt}
\usepackage{color}

\graphicspath{ {figs_vasp/} }

\allowdisplaybreaks

\begin{document}

\title{\vskip-3cm{\baselineskip14pt
    \begin{flushleft}
      \normalsize TTP19-001
    \end{flushleft}} \vskip1.5cm 
  Four-loop quark form factor 
  with quartic fundamental colour factor
  }

\author{
  Roman N. Lee$^{a}$,
  Alexander V. Smirnov$^{b}$,
  \\
  Vladimir A. Smirnov$^{c,d}$,
  Matthias Steinhauser$^{d}$,
  \\[1em]
  {\small\it (a) Budker Institute of Nuclear Physics,}\\
  {\small\it  630090 Novosibirsk, Russia}
  \\
  {\small\it (b) Research Computing Center, Moscow State University}\\
  {\small\it 119991, Moscow, Russia}
  \\  
  {\small\it (c) Skobeltsyn Institute of Nuclear Physics of Moscow State University}\\
  {\small\it 119991, Moscow, Russia}
  \\
  {\small\it (d) Institut f{\"u}r Theoretische Teilchenphysik,
    Karlsruhe Institute of Technology (KIT)}\\
  {\small\it 76128 Karlsruhe, Germany}  
}
  
\date{}

\maketitle

\thispagestyle{empty}

\begin{abstract}

  We analytically compute the four-loop QCD corrections for the colour
  structure $(d_F^{abcd})^2$ to the massless non-singlet quark form factor.
  The computation involves non-trivial non-planar integral families which have
  master integrals in the top sector. We compute the master integrals by
  introducing a second mass scale and solving differential equations with
  respect to the ratio of the two scales. We present details of our
  calculational procedure. Analytical results for the cusp and collinear
  anomalous dimensions, and the finite part of the form factor are
  presented. We also provide analytic results for all master integrals
  expanded up to weight eight.

\end{abstract}

\thispagestyle{empty}


\newpage


\section{Introduction}

Form factors are indispensable vertex functions which enter a number of
quantities in precision physics. Most prominent examples are the virtual
corrections to the Drell-Yan process or inclusive Higgs boson
production. Form factors are furthermore the simplest Green's function with a
non-trivial infrared structure. In fact, from the pole parts of the form
factors it is possible to extract universal quantities, like the cusp or
collinear anomalous dimension. They enter general formulae which predict the
infrared pole structure of massless on-shell multi-loop multi-leg QCD
amplitudes~\cite{Gardi:2009qi,Becher:2009qa}.

In this paper we consider the quark-anti-quark-photon form factor
with massless quarks which is obtained from the corresponding vertex
function $\Gamma^\mu_q$ via
\begin{eqnarray}
  F_q(q^2) &=& -\frac{1}{4(1-\epsilon)q^2}
  \mbox{Tr}\left( q_2\!\!\!\!\!/\,\,\, \Gamma^\mu_q \, q_1\!\!\!\!\!/\,\,\,
  \gamma_\mu\right)
               \label{eq::Fq}
  \,,
\end{eqnarray}
where we work in $d=4-2\epsilon$ space-time dimensions, $q=q_1+q_2$, and $q_1$
($q_2$) is the incoming quark (anti-quark) momentum.

Two-loop corrections to $F_q$ have been computed for the first time more than
twenty years
ago~\cite{Kramer:1986sg,Matsuura:1987wt,Matsuura:1988sm,Gehrmann:2005pd} and
the three-loop terms are available since about ten
years~\cite{Baikov:2009bg,Gehrmann:2010ue,Lee:2010ik,Gehrmann:2010tu} (for the
computation of master integrals see also Ref.~\cite{Heinrich:2009be}).  Only
two years ago first four-loop result for $F_q$ became available: In a first
step the large-$N_c$ limit has been considered, where only planar Feynman
diagrams contribute, and the fermionic and non-fermionic corrections have been
computed in Refs.~\cite{Henn:2016men} and~\cite{Lee:2016ixa},
respectively. Fermionic corrections with three closed quark loops have been
computed in Ref.~\cite{vonManteuffel:2016xki}; the complete terms proportional
to $n_f^2$ are available from~\cite{Lee:2017mip}.

Important information about QCD amplitudes is already obtained from the pole
part of the form factor. Of particular interest in this respect is the cusp
anomalous dimension, $\gamma_{\rm cusp}$~\cite{Korchemsky:1987wg}, which can
be extracted from the $1/\epsilon^2$ pole of $F_q$.  At three-loop order first
results for $\gamma_{\rm cusp}$ have been computed from the asymptotic
behaviour of splitting functions~\cite{Moch:2004pa} where the
fractional hadron momentum tends to 1. The results have been confirmed
afterwards by a dedicated calculation of the pole parts of the form
factor~\cite{Moch:2005id}. Also
at four-loop order there are two approaches to obtain $\gamma_{\rm cusp}$: The
$n_f^3$ terms of $\gamma_{\text{cusp}}$ has been obtained in
Refs.~\cite{Gracey:1994nn,Beneke:1995pq,vonManteuffel:2016xki} and analytic
results in the large-$N_c$ limit and for the (complete) $n_f^2$ contributions
have been obtained in Refs.~\cite{Henn:2016men,Lee:2016ixa,Lee:2017mip}
and~\cite{Davies:2016jie} from the explicit calculation of the form factor and
the splitting functions in the threshold limit, respectively. The approach
used in~\cite{Davies:2016jie} could be extended to all colour structures;
numerical results are presented in Refs.~\cite{Moch:2017uml,Moch:2018wjh}.
Recently the abelian four-loop contribution of the linear $n_f$ term to
$\gamma_{\text{cusp}}$ has been computed analytically in
Ref.~\cite{Grozin:2018vdn}. The main focus of~\cite{Grozin:2018vdn} is the cusp
anomalous dimension for massive fermions in QED. The abelian $n_f$
term for massless quarks is obtained as a by-product.

We define the expansion
of $F_q$ in terms of the bare strong coupling constant as
\begin{eqnarray}
  F_q &=& 1 +
  \sum_{n\ge1} 
  \left(\frac{\alpha_s^0}{4\pi}\right)^n
  \left(\frac{\mu^2}{-q^2-i0} \right)^{n\epsilon}
  F_q^{(n)}
  \,,
  \label{eq::FFbare}
\end{eqnarray}
The universal quantities $\gamma_{\rm cusp}$ and $\gamma_q$ are conveniently
extracted from the pole part of $\log(F_q)$ after renormalization of
$\alpha_s$ (see, e.g.,
Refs.~\cite{Korchemsky:1987wg,Becher:2009qa,Gehrmann:2010ue}). We define the
corresponding $n$-loop coefficients as follows
\begin{eqnarray}
  \gamma_x &=& \sum_{n\ge0} \left(\frac{\alpha_s{(\mu^2)}}{4\pi}\right)^n
  \gamma_x^n
  \,,
  \label{eq::gamma_x}
\end{eqnarray}
with $x=\text{cusp}$ or $x=q$. In order to fix the normalization we provide
the one-loop results which read $\gamma_{\rm cusp}^0 =4$ and $\gamma_{q}^0 =
-3 C_F$ (with $C_F=(N_c^2-1)/(2N_c)$).

\begin{figure}[t] 
  \begin{center}
    \begin{tabular}{ccc}
      \includegraphics[width=0.2\textwidth]{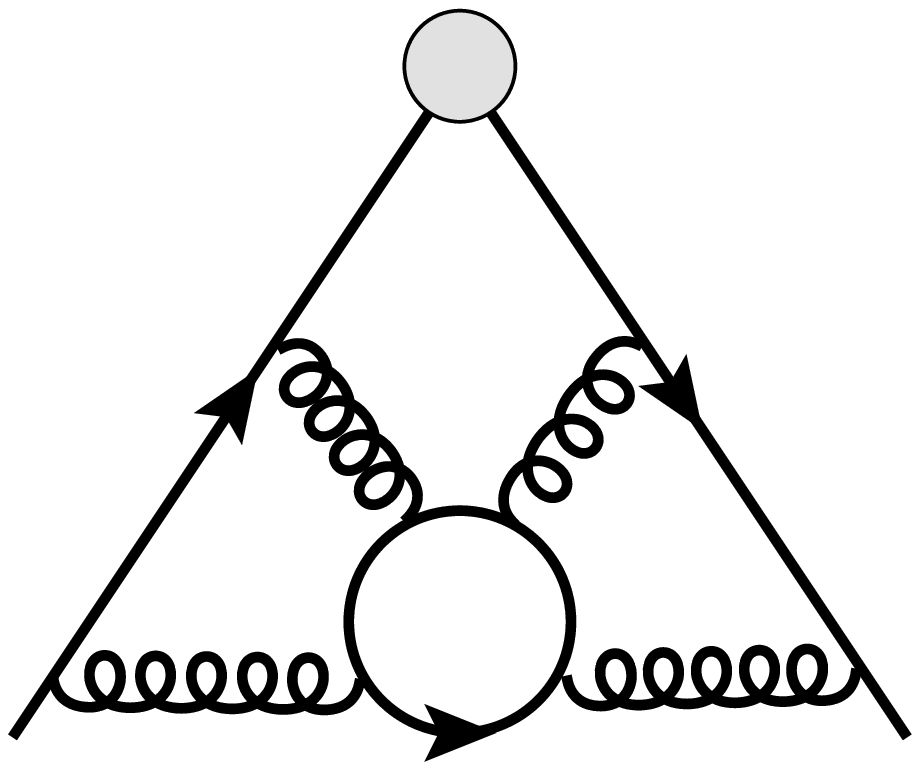} 
      & \hspace*{2em} &
      \includegraphics[width=0.2\textwidth]{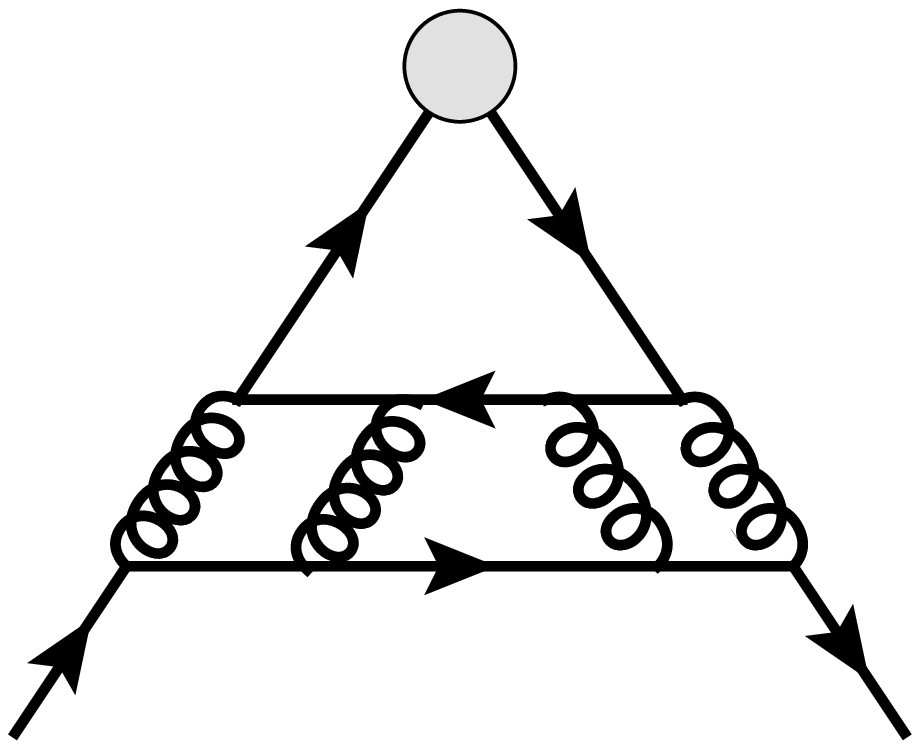}
      \\ (a) & & (b)
    \end{tabular}
    \caption{\label{fig::diag_df2}Non-singlet (a) and singlet (b) sample
      diagrams contributing to the colour structure $(d_F^{abcd})^2$ of the
      photon-quark form factor.  The gray blob indicates the external vector
      current.}
  \end{center}
\end{figure}

In this work, we provide analytic four-loop results for $\gamma_{\rm cusp}$,
$\gamma_q$ and $F_q$ for the colour structure
$(d_F^{abcd})^2$ which for a SU$(N_c)$ group is given by
\begin{eqnarray}
  \frac{(d_F^{abcd})^2}{N_A} 
  &=& \frac{N_c^4 - 6 N_c^2 + 18}{96 N_c^2}
      \,,
      \label{eq::d44}
\end{eqnarray}
with $N_A=N_c^2-1$.  Such colour factors arise from diagrams where four gluons
connect the two external fermion lines, see
Fig.~\ref{fig::diag_df2}(a). Note that there
are also singlet diagrams with colour factor proportional to $(d_F^{abcd})^2$,
see Fig.~\ref{fig::diag_df2}(b). In this work we only consider non-singlet
contributions.

\section{Calculation}

There are 18 Feynman diagrams with a closed fermion loop which is connected to
the external fermion line via four gluons. A representative diagram is shown
in Fig.~\ref{fig::diag_df2}(a); all other diagrams are obtained by the various
possibilities to connect the four gluons to the external fermion lines.

\begin{figure}[t] 
  \begin{center}
    \begin{tabular}{cccc}
      &
      \includegraphics[width=0.2\textwidth]{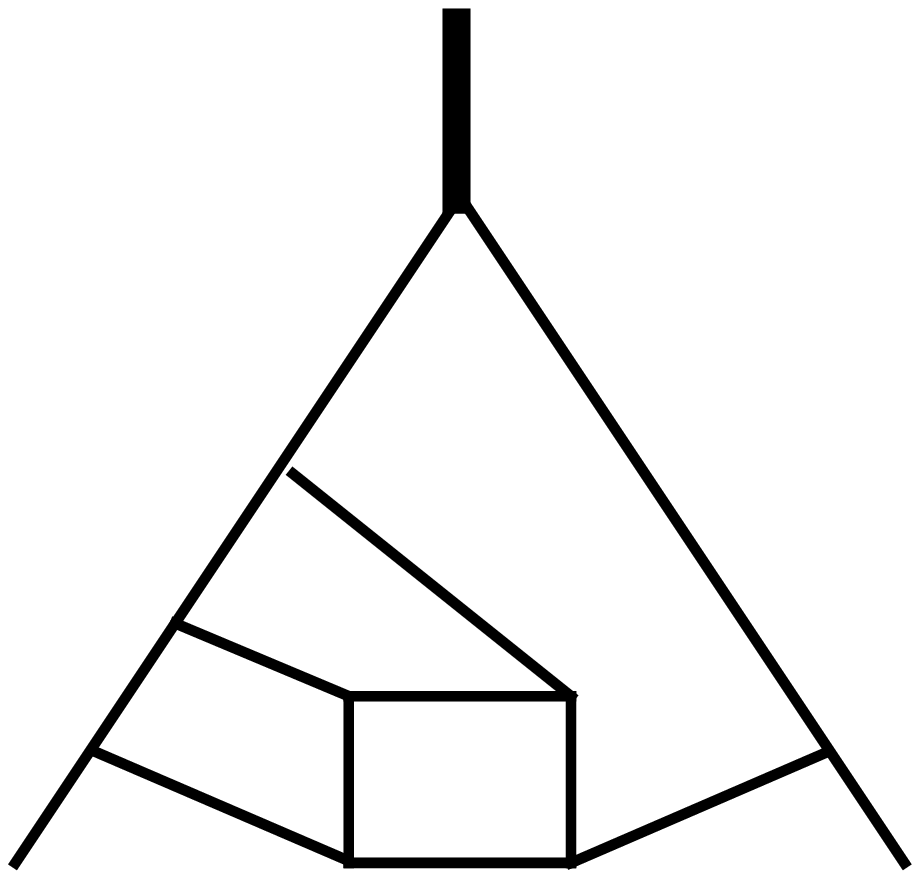} &
      \includegraphics[width=0.2\textwidth]{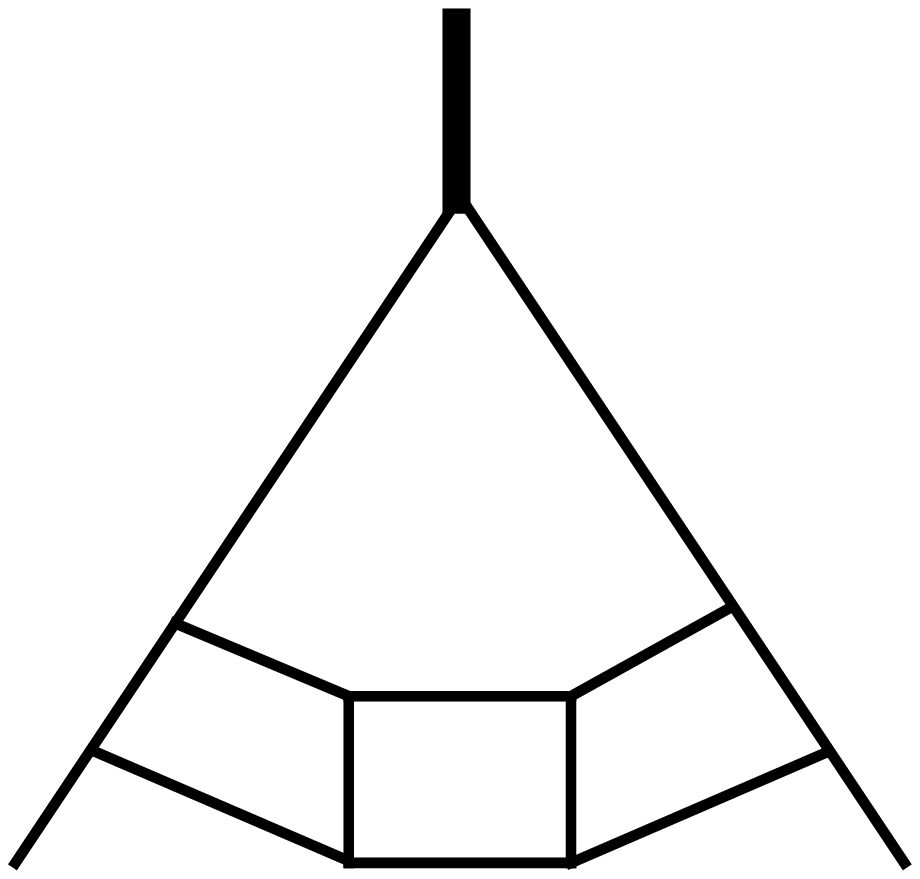} \\
      & 6 & 25 \\
      \includegraphics[width=0.2\textwidth]{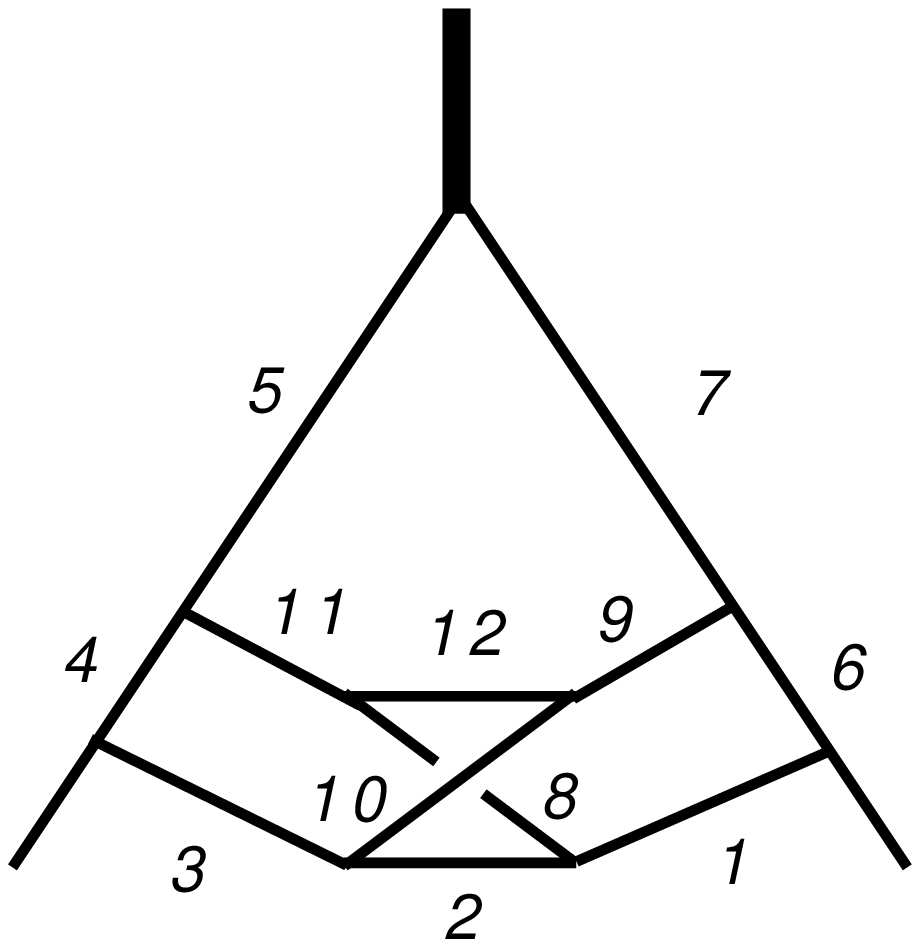} &
      \includegraphics[width=0.2\textwidth]{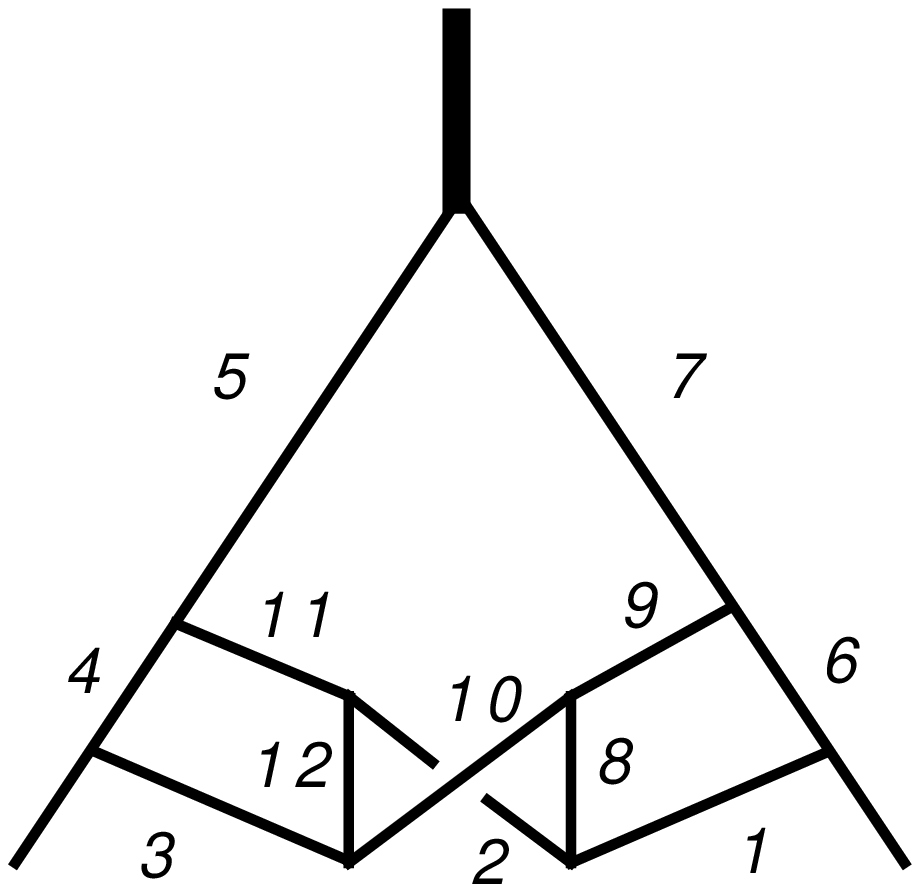} &
      \includegraphics[width=0.2\textwidth]{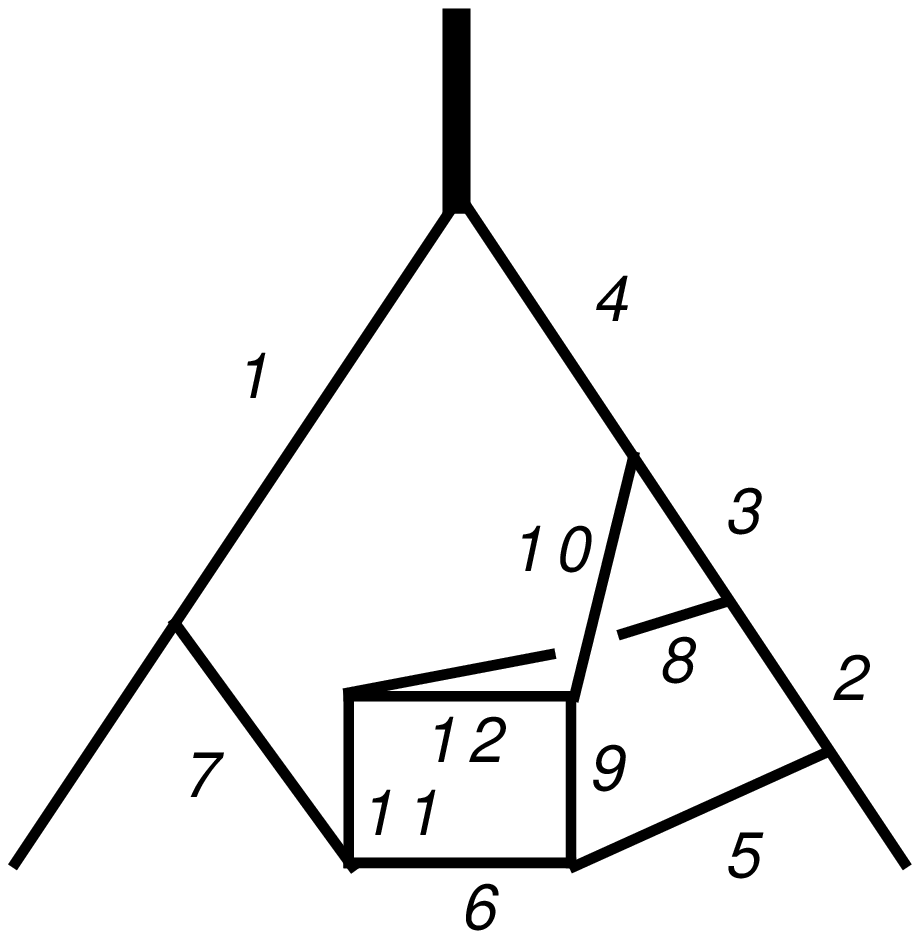} &
      \includegraphics[width=0.2\textwidth]{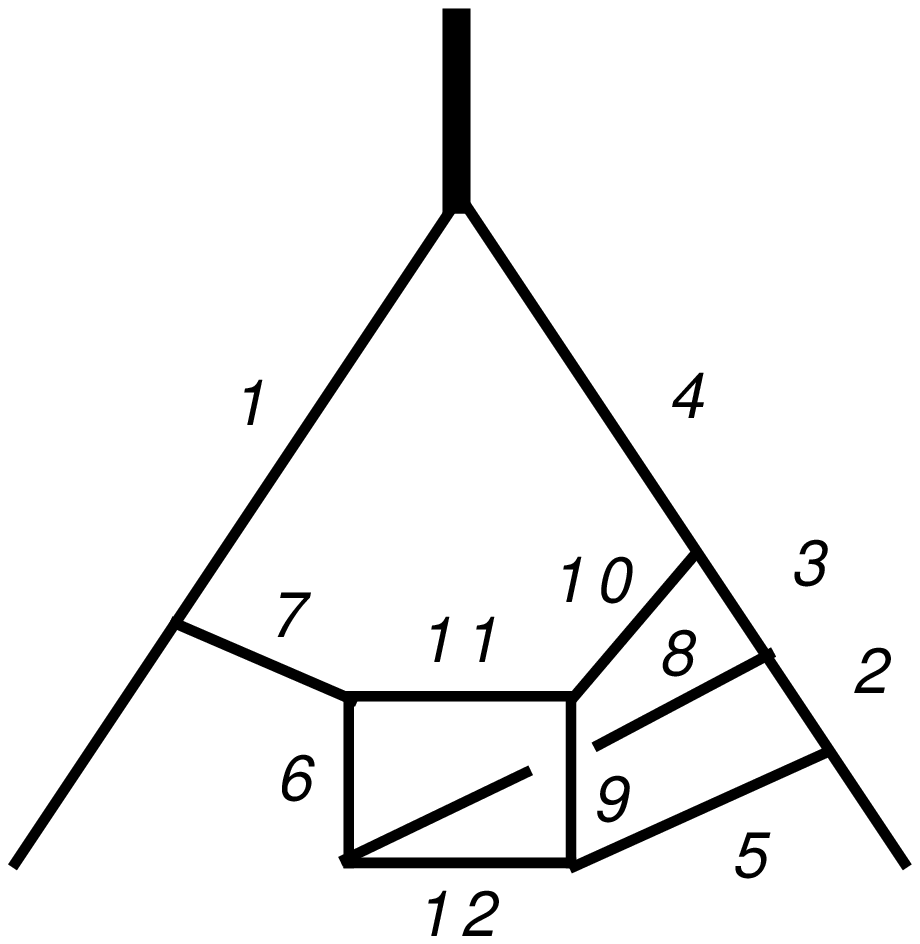} \\
      df2-2 & df2-3 & df2-5 & df2-6 \\
    \end{tabular}
    \caption{\label{fig::fam_df2}Planar (top row) and non-planar (bottom row)
      integral families. The numbers $n$ next to the lines correspond to the
      indices of the propagators, i.e. to the $n^{\rm th}$ integer argument of
      the functions representing the integrals. In addition to the 12
      propagators we have for each family six linear independent numerator
      factors. However, the corresponding indices are always zero for our
      master integrals.}
  \end{center}
\end{figure}

We can map the 18 (six planar and twelve non-planar) diagrams to six integral
families, two planar and four non-planar ones. They are illustrated in
Fig.~\ref{fig::fam_df2} where thin solid lines represent massless
propagators.\footnote{For convenience we use the internal numeration of the
  families also in the paper.} The thick external line carries the virtuality
$q^2$.  The planar families have been studied in
Refs.~\cite{Henn:2016men,Lee:2016ixa} where in particular all master integrals
have been computed. Results for the non-planar families in
Fig.~\ref{fig::fam_df2} are not yet available in the literature. In the
following we concentrate our discussion on them.

With the help of a suitably chosen projector to obtain $F_q$ (introduced in
Eq.~(\ref{eq::Fq})) we can express the amplitude as a linear combination of
scalar functions, which correspond to the family definitions of
Fig.~\ref{fig::fam_df2}.  All of them have 18 indices each, twelve for the
propagators and six for irreducible numerators.  We use {\tt
  FIRE}~\cite{Smirnov:2008iw,Smirnov:2013dia,Smirnov:2014hma} in combination
with {\tt LiteRed}~\cite{Lee:2012cn,Lee:2013mka} for the reduction to master
integrals. In Tab.~\ref{tab::df2} we present some information about the
individual (non-planar) families. Altogether we have to compute about 50\,000
integrals which can be reduced to almost 200 master integrals.  We refrain
from minimizing the master integrals among the various families since our
approach (see below) is applied to a whole family and provides simultaneous
results for all master integrals. We nevertheless establish relations between
master integrals of different families and use them as cross checks for
our results. For example, 36 of the 41 master integrals from df2-5 can be mapped
to master integrals of df2-2.
Note that we have performed the calculation in Feynman gauge.

\begin{table}[t]
  \begin{center}
    \begin{tabular}{c|c|c|c|c}
      non-planar   & \# 1-scale & \# 2-scale & number of & size of tables  \\
      family       & MIs  & MIs  & integrals & (MB) (1-scale)   \\
      \hline
      df2-2 &  71 & 337 & 14156 & 98  \\
      df2-3 &  45 & 244 & 15278 & 50  \\
      df2-5 &  41 &  92 & 11620 & 23  \\
      df2-6 &  35 &  78 & 11531 & 18  \\
    \end{tabular}
    \caption{\label{tab::df2}Information about the non-planar families.}
  \end{center}
\end{table}

For the computation of the master integrals we use the idea suggested
in~\cite{Henn:2013nsa} and used in our previous works for the
planar~\cite{Henn:2016men,Lee:2016ixa} and $n_f^2$
calculation~\cite{Lee:2017mip}: we introduce a second mass scale
$q_2^2 = x q^2$ as the virtuality of one of the external quarks. This
increases, of course, the complexity of the problem. We encounter a more
difficult reduction problem and there are significant more master integrals
present in the individual families (compare ``\# 1-scale MIs''
  and ``\# 2-scale MIs'' in Tab.~\ref{tab::df2}).  However, the
introduction of the second mass scale has the advantage that we can use the
powerful method of differential equations.  In fact, the basic idea is to
choose $x=1$ in order to fix the boundary conditions, since in this limit one
has to compute massless two-point functions which are well studied in the
literature~\cite{Baikov:2010hf,Lee:2011jt}. The differential equations are
then used to transport the information to the point $x=0$.

The method has been described in some details in Ref.~\cite{Lee:2017mip} where
for the first time non-planar four-loop families have been considered. For the
integral families considered in this paper the method had to be further
refined. Note that in Ref.~\cite{Lee:2017mip} no non-planar master integrals
had to be computed in the top sector where the indices of all twelve
propagators are positive.

For each family we can introduce a system of differential equations
of the form\footnote{In the following we do not explicitly show the
  $\epsilon$ dependence of the functions in the arguments.}
\begin{eqnarray}
  \partial_x j(x) &=& m(x)  j(x)\,,
                   \label{eq::j}
\end{eqnarray}
where $j(x)$ is a vector of (two-scale) master integrals in the primary basis
chosen by {\tt FIRE} and $m(x)$ is a square matrix.  We use the idea suggested in
Refs.~\cite{Henn:2013pwa,Henn:2014qga} to turn to a so-called $\epsilon$ or
canonical basis where the right-hand side of the differential equations is
proportional to $\epsilon$ and singularities with respect to the variables of
the differential equations are Fuchsian, i.e., of the form $1/(x-a)$.  To
arrive at a canonical basis, we use the algorithm of
Ref.~\cite{Lee:2014ioa}\footnote{Meanwhile there are two public computer
  implementations of this algorithm, see
  Refs.~\cite{Gituliar:2016vfa,Gituliar:2017vzm,Meyer:2016slj,Prausa:2017ltv}. A
  somewhat different approach to the same problem can be found in
  Ref.~\cite{Meyer:2017joq}.}  and its private implementation. We apply this
procedure to each family separately and arrive at an $\epsilon$ form given by
\begin{eqnarray}
  \partial_x J(x) &=& \epsilon M(x) J(x)\,,
                   \label{eq::J}
\end{eqnarray}
where $J$ are the master integrals in the canonical basis, which are connected
to the ones in the primary basis via $j(x)=T(x) J(x)$. The matrix $M(x)$ only has a
simple dependence on $x$
\begin{eqnarray}
  M(x) &=& \sum_a \frac{M_a}{x-a}\,,
           \label{eq::M}
\end{eqnarray}
with constant matrices $M_a$. In our case the sum only includes two terms,
$a=0$ and $a=1$, which correspond to the physical point and the point
where we want to fix the boundary conditions, respectively.
Next, we introduce, as in~\cite{Lee:2017mip}, the path-ordered
exponent
\begin{eqnarray}\label{eq:pexp}
  U(x,x_0) &=& P\exp\left[\epsilon\int\limits_{x_0}^{x} {\rm d}\xi M(\xi)\right]\,,
\end{eqnarray}
and define the quantities (with a slight abuse of notation)\footnote{Note that
  $U(x,0)$ and $U(x,1)$ as
  defined in Eq.~\eqref{eq:pexp} are divergent and thus confusion with
  Eq.~(\ref{eq::Ux01}) is excluded.}
\begin{eqnarray}
  U(x,0) &=& \lim_{x_0\to 0} U(x,x_0) x_0^{\epsilon A_0}\,,\nonumber\\
  U(x,1) &=& \lim_{x_0\to 1} U(x,x_0) (1-x_0)^{\epsilon A_1}\,,
             \label{eq::Ux01}
\end{eqnarray}
which have the properties
\begin{eqnarray}
  U(x,0) &\stackrel{x\to 0}{\longrightarrow}& x^{\epsilon A_0}\,,\nonumber\\
  U(x,1) &\stackrel{x\to 1}{\longrightarrow}& (1-x)^{\epsilon A_1}\,.
\end{eqnarray}
Note that $U(x,0)$ and $U(x,1)$ can be obtained in a straightforward way as an
expansion in $\epsilon$ in terms of Harmonic
polylogarithms (HPLs)~\cite{Remiddi:1999ew} with arguments $(1-x)$ and $x$,
respectively.  Furthermore, both $U(x,0)$ and $U(x,1)$ solve the
system~(\ref{eq::M}) and are thus related by a matrix $U_{01}$ which only
depends on $\epsilon$ but not on $x$:
\begin{eqnarray}
  U(x,1) &=& U(x,0) U_{01} \,.
             \label{eq::U01}
\end{eqnarray}
We will call the matrix $U_{01}\equiv U_{01}(\epsilon)$ the {\it associator}. It can be constructed
by multiplying Eq.~(\ref{eq::U01}) by $x^{-\epsilon A_0}$ from the left
and taking the limit $x\to0$ which leads to
\begin{eqnarray}
  U_{01} &=& \lim_{x\to0} x^{-\epsilon A_0} U(x,1) \,.
             \label{eq::U01_2}
\end{eqnarray}
In practice, the right-hand side of Eq.~(\ref{eq::U01_2}) is evaluated by
extracting all $\log(x)$ terms contained in $U(x,1)$ with the help
of shuffle relations to eliminate the leading letter ``1'' from the
HPLs.\footnote{Note that the program package {\tt HPL}~\cite{Maitre:2005uu}
  has a build-in command which can be used for this step.}  They have to
cancel against the $\log(x)$ terms from $x^{-\epsilon A_0}$ such that the
limit $x\to 0$ can be taken.

Let us in a next step discuss the boundary conditions which we compute for
$x=1$. Note that in this limit our integrals are analytical and thus we do not
have contributions of the form $x^{-k\epsilon}$ with $k\not=0$.
In the canonical basis we can thus write
\begin{eqnarray}
  J(x) & = & U(x,1)C_1 \,,
            \label{eq::J1}
\end{eqnarray}
where $C_1$ is a vector with $\epsilon$-dependent components. Similarly we
have
\begin{eqnarray}
  J(x) & = & U(x,0)C_0 \,.
            \label{eq::J0}
\end{eqnarray}
Note that in this limit the integrals in $J$ have a logarithmic dependence on
$x$. We are only interested in the so-called hard part which means that from
the various contributions of the form $x^{-k\epsilon}$ we only take those with
$k=0$.

Next we want to relate the constants $C_0$ and $C_1$ to coefficients of
integrals from the primary basis evaluated near $x=0$ and $x=1$,
respectively. These relations have the form
\begin{eqnarray}
  C_0 & = & L_0 c_0\,, \nonumber\\
  C_1 & = & L_1 c_1\,,
            \label{eq::L01}
\end{eqnarray}
where $L_{0,1}$ are matrices depending on $\epsilon$, and $c_{0,1}$ are the
column vectors of the specific coefficients in the asymptotics $x\to 0$ and $x\to 1$,
respectively. Note that the vector $c_1$ is obtained from the boundary
conditions, and the aim of our calculation is the hard part of $c_0$.  In the
following we present details about how we determine which set of coefficients $c_0$
suffices and calculate the matrix $L_0$. $L_{1}$ and $c_{1}$ are calculated
in analogy.

We start with the generalized series expansion of
$T\left(x,\epsilon\right)U\left(x,{0}\right)$ which can be cast in the form
\begin{eqnarray}
  T\left(x,\epsilon\right)U\left(x,{0}\right)
  &=&\sum_{\alpha,k}u\left(\alpha,k\right)x^{\alpha}\log^{k}x\,,
      \label{eq::TU}
\end{eqnarray}
where $\alpha = n_1 + \epsilon n_2$ with integer $n_1$ and $n_2$, and
$u\left(\alpha,k\right)$ are matrices which depend on $\epsilon$.  The key
point is that, using the approach of Ref.~\cite{Lee:2017qql}, we can calculate
plenty of terms in the above {expression, keeping the exact $\epsilon$
  dependence}.  After applying Eq.~(\ref{eq::TU}) to $C_0$ we have
\begin{eqnarray}
  j(x)&=&\sum_{\alpha,k}c\left(\alpha,k\right)x^{\alpha}\log^{k}x\,,
\end{eqnarray}
where
\begin{eqnarray}
  c\left(\alpha,k\right)&=&u\left(\alpha,k\right)C_{0}\,.
\end{eqnarray}
Each $c\left(\alpha,k\right)$ is a column vector of the form
$\left(c_{1}\left(\alpha,k\right),\ldots,c_{N}\left(\alpha,k\right)
\right)^{\intercal}$, where $N$ is the number of two-scale master integrals
of the considered family.

In a next step we select from the coefficients $c_{i}\left(\alpha,k\right)$
(for various $i$, $\alpha$, and $k$) the minimal set, which is sufficient to
determine all constants in
$C_{0}=\left(C_{01},\ldots C_{0N}\right)^{\intercal}$.  Let this set be
\begin{align}
c_{i_{1}}\left(\alpha_{1},k_{1}\right) & =\sum_{j=1}^{N}u_{i_{1}j}\left(\alpha_{1},k_{1}\right)C_{0j},\nonumber\\
& \vdots\nonumber\\
c_{i_{M}}\left(\alpha_{M},k_{M}\right) & =\sum_{j=1}^{N}u_{i_{M}j}\left(\alpha_{M},k_{M}\right)C_{0j}\,.
\end{align}
Here \emph{sufficient} refers to the rank of the matrix
\begin{eqnarray}
  R&=&\begin{bmatrix}u_{i_{1}1}
    \left(\alpha_{1},k_{1}\right) & \ldots & u_{i_{1}N}
    \left(\alpha_{1},k_{1}\right)\\
    \vdots & \ddots & \vdots\\
    u_{i_{M}1}\left(\alpha_{M},k_{M}\right) & 
    \ldots & u_{i_{M}N}\left(\alpha_{M},k_{M}\right)
  \end{bmatrix}
\end{eqnarray}
which has to be greater or equal to the number of master integrals $N$,
and \emph{minimal} means that $M=N$. In
other words, $R$ is a square matrix, which is invertible and we have
\begin{align}
  c_{0} & =\left(c_{i_{1}}\left(\alpha_{1},k_{1}\right),\ldots,c_{i_{N}}\left(\alpha_{N},k_{N}\right)\right)^{\intercal}\,,\nonumber\\
  L_{0} & =R^{-1}.
\end{align}
Of course, this procedure does not lead to unique quantities $c_0$ and $L_0$,
which, however, is not a problem since the arbitrariness cancels after
performing the matching to the one-scale master integrals. As a rule of
thumb we first try to pick coefficients only among the leading coefficients of
the asymptotic expansion of the integrals $j(x)$ and then extend the search to
subleading terms in $x$, if necessary.

Using Eqs.~(\ref{eq::U01}),~(\ref{eq::J1}),~(\ref{eq::J0})
and,~(\ref{eq::L01}) we finally arrive at
\begin{eqnarray}
  c_0 &=& L_0^{-1} U_{01} L_1 c_1 \,,
\end{eqnarray}
which is used to obtain the coefficients at $x=0$ from the ones at $x=1$.
Note that $L_0$ and $L_1$ are exact in $\epsilon$ but $U_{01}$ is usually
known as an expansion for $\epsilon\to 0$.

The number of components of $c_0$ is the number of the two-scale master
integrals. For example, for df2-2, it is 337. Our goal is the determination of
the coefficients in the naive part of the expansion, i.e. the part of the
expansion with non-negative integer powers of $x$.  For df2-2, $c_0$ contains
116 coefficients corresponding to the naive limit. One can expect that this
number is equal to the number of one-scale master integrals, 
  which is, however, not the case. The reason is the additional symmetry of the one-scale integrals,
related to the permutation of two massless legs. This symmetry reduces the
number of one-scale master integrals to 71. Therefore, there are $116-71 = 45$
redundant relations which we use as a check once we have satisfied 71
relations using explicit results for the one-scale master integrals. In
practice, most of the one-scale master integrals have the same indices as the
corresponding two-scale master integrals so that the results for these
one-scale master integrals are obtained directly from the naive part of the
two-scale master integrals.  For the remaining one-scale master integrals
(where an index equal to two is chosen in another place), results are obtained
after solving simple linear systems of equations.

Let us stress that the basic ideas of the described procedure have already
been discussed in Ref.~\cite{Lee:2017mip}, However, the approach presented
here is more algorithmic and has now reached a state where it can be applied
to highly non-trivial non-planar integral families, as it is demonstrated in
this paper.

Note that in our case, we had to expand $U_{01}$ up to $\epsilon^9$ (weight 9)
for df2-2 and df2-3 since the property of uniform transcendentality is
destroyed when mapping the two-scale master integrals to one-scale master
integrals in the limit $x\to 0$. In the final result for the form factor all
weight-nine constants drop out. This happens separately for each family.  In
principle it is possible to adapt the basis of the one-scale master integrals
such that only an expansion of $U_{01}$ up to $\epsilon^8$ is
necessary. However, our approach is powerful enough such that an
expansion up to $\epsilon^9$ did not pose any serious technical problems.  For
df2-5 and df2-6 an expansion up to weight eight is sufficient.

The reduction of one-scale as well as of two-scale integrals, needed
for the derivation of differential equations for the (two-scale) master
integrals, took several months for each of the four non-planar families. Using
the standard version of {\tt FIRE} we have failed to reduce the two-scale
integrals of family df2-2 in the top sector. However, following the ideas of
Ref.~\cite{vonManteuffel:2016xki}, based on modular arithmetics, we managed to
improve the performance of {\tt FIRE}~\cite{smirnov_FIRE}. The new version can
be used in a massive parallel mode on supercomputers which allows us 
to obtain the missing reductions.

In Ref.~\cite{Boels:2017ftb} many (planar and non-planar) four-loop vertex
integrals have been computed numerically. Among them are uniformly
trancendental integrals in the top sectors of df2-2 and df2-3. Reducing these
integrals to our primary bases and using our analytic results we can confirm
the results (A.4)--(A.7) of Ref.~\cite{Boels:2017ftb}.
  
Let us finally mention that we have performed numerical cross checks of all
master integrals of families df2-2, df2-3, df2-5 and df2-6 with up to ten
positive indices expanded up to order $\epsilon^0$ using {\rm
  FIESTA}~\cite{Smirnov:2015mct}.

Analytic results for all master integrals can be downloaded in electronic form
from~\cite{progdata}. For illustration we show for families df2-2 and df2-3
the master integrals with twelve lines in the Appendix.
Families df2-5 and df2-6 have no twelve-line master integrals.


\section{Results}

After inserting the analytic results for the master integrals into the
amplitude for the form factor we observe that all poles higher than $1/\epsilon^2$
cancel. This is expected since the coefficients of the
$1/\epsilon^8, \ldots, 1/\epsilon^3$ poles are determined by lower-loop
contributions. Since the colour structure
$(d_F^{abcd})^2$ appears for the first time at four-loop order it can at most
have $1/\epsilon^2$ poles. For the same reason there are no renormalization
contributions to the $(d_F^{abcd})^2$ contribution.

Our result for $F_q^{(4)}$ (see Eq.~(\ref{eq::FFbare})) reads
\begin{eqnarray}
  F_q^{(n)}\Big|_{(d_F^{abcd})^2} 
  &=& 
      n_f \frac{(d_F^{abcd})^2}{N_F} \Bigg\{
\frac{1}{\epsilon^2}  \Bigg[
\frac{40 \zeta _5}{3}
+\frac{8 \zeta _3}{3}
-\frac{4 \pi ^2}{3}
\Bigg]
+\frac{1}{\epsilon}  \Bigg[
-\frac{148 \pi ^6}{8505}
-\frac{152 \zeta _3^2}{3}-\frac{8 \pi ^2 \zeta _3}{3}
\nonumber\\&&\mbox{}
+\frac{2720 \zeta _5}{9}
+\frac{10  \pi ^4}{27}
+\frac{664 \zeta _3}{9}
-\frac{284 \pi ^2}{9}
+48
\Bigg]
-1240 \zeta _7
-\frac{988 \pi ^4 \zeta _3}{135}
\nonumber\\&&\mbox{}
+\frac{496 \pi ^2 \zeta      _5}{9}
+\frac{10405 \pi ^6}{10206}
+ \frac{680 \zeta _3^2}{9}
+\frac{95098 \zeta _5}{27}
+\frac{46 \pi ^2 \zeta    _3}{9}
+\frac{1888 \pi ^4}{405}
\nonumber\\&&\mbox{}
-\frac{13414 \zeta _3}{27}
-\frac{10783 \pi ^2}{27}
+\frac{3190}{3}
      \Bigg\}
              \,,
              \label{eq::Fq_res}
\end{eqnarray}
where $N_F=N_c=3$ and $\zeta_n$ is Riemann's zeta function evaluated at
$n$.

The cusp and collinear anomalous dimension can be extracted from the
$1/\epsilon^2$ and $1/\epsilon$ poles, respectively. For convenience of the
reader we present the corresponding results separately. They are given by
\begin{eqnarray}
  C_F \gamma_{\rm cusp}^3\Big|_{(d_F^{abcd})^2} 
  &=&
      n_f \frac{(d_F^{abcd})^2}{N_F} \Bigg(
      - \frac{1280}{3}\zeta_5
      - \frac{256}{3}\zeta_3
      + \frac{128}{3}\pi^2
      \Bigg)
      \nonumber\\
  &\approx& n_f \frac{(d_F^{abcd})^2}{N_F} \left( -123.894910\ldots \right)
            \,,
  \label{eq::cusp}\\
  \gamma_{q}^3\Big|_{(d_F^{abcd})^2} 
  &=&
      n_f \frac{(d_F^{abcd})^2}{N_F} \Bigg(
      -\frac{592 \pi ^6}{8505}
      -\frac{608 \zeta _3^2}{3}
      +\frac{10880 \zeta _5}{9}
      -\frac{32 \pi ^2 \zeta _3}{3}
      \nonumber\\&&\mbox{}
      +\frac{40 \pi ^4}{27}
      +\frac{2656 \zeta _3}{9}
      -\frac{1136 \pi ^2}{9}
      +192
                    \Bigg)
                    \,.
  \label{eq::q}
\end{eqnarray}

In Refs.~\cite{Moch:2017uml,Moch:2018wjh} the quark and gluon
splitting functions at four-loop order have been considered. As a by-product
numerical results for cusp anomalous dimensions have been obtained, in
particular for $C_F \gamma_{\rm cusp}^3|_{(d_F^{abcd})^2}$ as given in
Eq.~(\ref{eq::cusp}). The numerical result from Tab.~2 of~\cite{Moch:2017uml}
reads $-123.90 \pm 0.2$ and agrees well with
the numerical evaluation of our analytic expression.

The results for $\gamma_{q}^3$ and the finite part of the form factor
in Eqs.~(\ref{eq::q}) and~(\ref{eq::Fq_res}) are new.


\section{\label{sec::con}Conclusions}

We perform the next step towards the computation of massless four-loop form
factors and compute the contribution of the quartic colour structure
$(d_F^{abcd})^2$ to the photon-quark form factor. We have to consider two
planar and four non-planar integral families which are shown in
Fig.~\ref{fig::fam_df2}.  We want to stress that this is the first time
that master integrals with twelve propagators corresponding to non-planar
graphs have to be considered.  Our main results are shown in
Eqs.~(\ref{eq::Fq_res}),~(\ref{eq::cusp}) and~(\ref{eq::q}). Furthermore, we
provide analytic results for all master integrals in a
supplementary file to this paper.

We have used this calculation to further refine our method, which is used to
obtain analytic results for the master integrals. The new element is the
construction of the so-called associator which directly relates the
coefficients in the boundary condition to the coefficients of the integrals
in the physical limit.  We are confident that the remaining contributions can
be computed along the same lines. However, one has to keep in mind that much
more families have to be considered and that the reductions to master integrals
(both with one and two mass scales) require a significant amount of CPU time.


\section*{\label{sec::ack}Acknowledgments}

The work of A.S. and V.S. is supported by RFBR, grant 17-02-00175.  The
work of R.L. is supported in part by RFBR grant 17-02-00830 and by ``Basis''
foundation for theoretical physics and mathematics.  We thank the High
Performance Computing Center Stuttgart (HLRS) for providing computing time.
The Feynman diagrams were drawn with the help of {\tt
  Axodraw}~\cite{Vermaseren:1994je} and {\tt JaxoDraw}~\cite{Binosi:2003yf}.


\begin{appendix}

\section*{Appendix: Explicit results for twelve-line non-planar master integrals}

In this appendix we present explicit results for the most complicated
master integrals of the families df2-2 and df2-3 with twelve lines.
We provide the $\epsilon$ expansion up to the constant term. Our results read
\begin{eqnarray}
\lefteqn{G_{111111111111}^{\text{(\rm df2-2)}}=}
\nonumber\\&&
+\frac{1}{\epsilon^8}  \Bigg[
\frac{1}{144}
\Bigg]
%
%
+\frac{1}{\epsilon^7}  \Bigg[
\frac{73}{576}
\Bigg]
%
%
+\frac{1}{\epsilon^6}  \Bigg[
\frac{331}{1152}-\frac{7 \pi ^2}{216}
\Bigg]
%
%
+\frac{1}{\epsilon^5}  \Bigg[
-\frac{311 \zeta _3}{216}-\frac{245 \pi ^2}{576}-\frac{1765}{1152}
\Bigg]
\nonumber\\&&\mbox{}
+\frac{1}{\epsilon^4}  \Bigg[
-\frac{1103 \zeta _3}{54}-\frac{37 \pi ^4}{1440}-\frac{917 \pi ^2}{1728}+\frac{2297}{576}
\Bigg]
%
%
+\frac{1}{\epsilon^3}  \Bigg[
\frac{4021 \pi ^2 \zeta _3}{648}-\frac{42053 \zeta _3}{1728}-\frac{22667 \zeta
               _5}{360}
\nonumber\\&&\mbox{}
-\frac{31327 \pi ^4}{51840}+\frac{2615 \pi ^2}{864}-\frac{59}{36}
\Bigg]
%
%
+\frac{1}{\epsilon^2}  \Bigg[
\frac{10784 \zeta _3^2}{81}+\frac{13595 \pi ^2 \zeta _3}{216}+\frac{293837
               \zeta _3}{1728}-\frac{268139 \zeta _5}{360}
\nonumber\\&&\mbox{}
-\frac{4901 \pi ^6}{38880}-\frac{40973 \pi ^4}{103680}-\frac{347 \pi ^2}{96}-\frac{21161}{288}
\Bigg]
%
%
+\frac{1}{\epsilon}  \Bigg[
\frac{1960259 \zeta _3^2}{1296}+\frac{1037 \pi ^4 \zeta _3}{160}+\frac{117521
               \pi ^2 \zeta _3}{1296}
\nonumber\\&&\mbox{}
-\frac{490831 \zeta _3}{864}+\frac{508661 \pi ^2 \zeta _5}{2160}-\frac{2028557
              \zeta _5}{2880}-\frac{10749139 \zeta _7}{4032}-\frac{3561371 \pi
              ^6}{2177280}+\frac{110171 \pi ^4}{34560}
\nonumber\\&&\mbox{}
-\frac{20797 \pi ^2}{432}+\frac{222407}{288}
\Bigg]
%
%
-\frac{4937 s_{8 a}}{6}-\frac{582209 \pi ^2 \zeta _3^2}{1944}+\frac{8605981
               \zeta _3^2}{5184}+\frac{2064401 \zeta _5 \zeta _3}{270}
\nonumber\\&&\mbox{}
+\frac{3543269 \pi ^4 \zeta _3}{77760}-\frac{876841 \pi ^2 \zeta
              _3}{1296}+\frac{325039 \zeta _3}{216}+\frac{87229 \pi ^2 \zeta
              _5}{48}+\frac{2528065 \zeta _5}{576}-\frac{8894555 \zeta
              _7}{504}
\nonumber\\&&\mbox{}
              -\frac{17509 \pi ^8}{1088640}+\frac{579329 \pi ^6}{2177280}-\frac{547763 \pi ^4}{51840}+\frac{126427 \pi ^2}{216}-\frac{1754951}{288}
%
%
+ {\cal O}(\epsilon)
\,,
\\
\lefteqn{G_{111111111112}^{\text{(\rm df2-2)}}=}
\nonumber\\&&
+\frac{1}{\epsilon^8}  \Bigg[
-\frac{1}{72}
\Bigg]
%
%
+\frac{1}{\epsilon^7}  \Bigg[
-\frac{83}{288}
\Bigg]
%
%
+\frac{1}{\epsilon^6}  \Bigg[
\frac{9163}{5184}+\frac{7 \pi ^2}{108}
\Bigg]
%
%
+\frac{1}{\epsilon^5}  \Bigg[
\frac{203 \zeta _3}{108}+\frac{857 \pi ^2}{864}+\frac{408031}{15552}
\Bigg]
\nonumber\\&&\mbox{}
+\frac{1}{\epsilon^4}  \Bigg[
\frac{7109 \zeta _3}{216}+\frac{59 \pi ^4}{720}-\frac{49285 \pi ^2}{7776}-\frac{101431}{1728}
\Bigg]
%
%
+\frac{1}{\epsilon^3}  \Bigg[
\frac{1703 \pi ^2 \zeta _3}{324}-\frac{3601769 \zeta _3}{7776}-\frac{8113
               \zeta _5}{180}
\nonumber\\&&\mbox{}
+\frac{15673 \pi ^4}{6480}-\frac{66752 \pi ^2}{729}-\frac{36323851}{139968}
\Bigg]
%
%
+\frac{1}{\epsilon^2}  \Bigg[
\frac{29089 \zeta _3^2}{162}+\frac{9137 \pi ^2 \zeta _3}{81}-\frac{119403089
               \zeta _3}{23328}
\nonumber\\&&\mbox{} 
-\frac{28579 \zeta _5}{120}-\frac{18167 \pi ^6}{68040}-\frac{1566377 \pi ^4}{155520}-\frac{1085407 \pi ^2}{11664}+\frac{13540370}{6561}
\Bigg]
%
%
+\frac{1}{\epsilon}  \Bigg[
\frac{2292335 \zeta _3^2}{648}
\nonumber\\&&\mbox{}
-\frac{50413 \pi ^4 \zeta _3}{2160}+\frac{2537869 \pi ^2 \zeta
              _3}{1458}+\frac{32307611 \zeta _3}{7776}+\frac{388549 \pi ^2
              \zeta _5}{1080}-\frac{155919821 \zeta _5}{12960}
\nonumber\\&&\mbox{}
-\frac{2985239 \zeta _7}{2016}-\frac{69407 \pi ^6}{38880}-\frac{466151 \pi ^4}{2880}+\frac{192951265 \pi ^2}{209952}-\frac{1928298269}{209952}
\Bigg]
\nonumber\\&&\mbox{}
%
-\frac{150569 s_{8 a}}{15}-\frac{801973}{972} \pi ^2 \zeta
              _3^2+\frac{1602372409 \zeta _3^2}{23328}-\frac{751148 \zeta _5
              \zeta _3}{135}-\frac{2514809 \pi ^4 \zeta _3}{9720}
\nonumber\\&&\mbox{} 
+\frac{412729031 \pi ^2 \zeta _3}{34992}+\frac{6680310761 \zeta
              _3}{209952}+\frac{7821953 \pi ^2 \zeta
              _5}{1080}-\frac{3525176537 \zeta _5}{38880}-\frac{101624527
              \zeta _7}{2016}
\nonumber\\&&\mbox{}
+\frac{62792629 \pi ^8}{27216000}-\frac{14403373 \pi ^6}{979776}+\frac{7488623 \pi ^4}{87480}-\frac{2098797893 \pi ^2}{629856}+\frac{33048481297}{944784}
%
\nonumber\\&&\mbox{}
+{\cal O}(\epsilon)\,,
\end{eqnarray}
\begin{eqnarray}
\lefteqn{G_{111111111111}^{\text{(\rm df2-3)}}=}
\nonumber\\&&
+\frac{1}{\epsilon^8}  \Bigg[
\frac{1}{144}
\Bigg]
%
%
+\frac{1}{\epsilon^7}  \Bigg[
\frac{5}{48}
\Bigg]
%
%
+\frac{1}{\epsilon^6}  \Bigg[
\frac{125}{576}-\frac{5 \pi ^2}{108}
\Bigg]
%
%
+\frac{1}{\epsilon^5}  \Bigg[
-\frac{401 \zeta _3}{216}-\frac{175 \pi ^2}{288}-\frac{235}{288}
\Bigg]
\nonumber\\&&\mbox{}
+\frac{1}{\epsilon^4}  \Bigg[
-\frac{1567 \zeta _3}{72}+\frac{19 \pi ^4}{576}-\frac{853 \pi ^2}{1728}+\frac{143}{64}
\Bigg]
%
%
+\frac{1}{\epsilon^3}  \Bigg[
\frac{13151 \pi ^2 \zeta _3}{1296}-\frac{13711 \zeta _3}{864}-\frac{16277
               \zeta _5}{360}
\nonumber\\&&\mbox{}
+\frac{5489 \pi ^4}{17280}+\frac{5905 \pi ^2}{1728}-\frac{289}{32}
\Bigg]
%
%
+\frac{1}{\epsilon^2}  \Bigg[
\frac{248513 \zeta _3^2}{1296}+\frac{40319 \pi ^2 \zeta _3}{432}+\frac{46481
               \zeta _3}{432}-\frac{3751 \zeta _5}{12}
\nonumber\\&&\mbox{}
+\frac{751 \pi ^6}{9720}-\frac{15833 \pi ^4}{103680}-\frac{21929 \pi ^2}{1728}+\frac{58997}{1152}
\Bigg]
%
%
+\frac{1}{\epsilon}  \Bigg[
\frac{388001 \zeta _3^2}{216}-\frac{653 \pi ^4 \zeta _3}{180}+\frac{731 \pi ^2
               \zeta _3}{2592}
\nonumber\\&&\mbox{}
-\frac{111755 \zeta _3}{288}+\frac{37751 \pi ^2 \zeta _5}{216}+\frac{26203
              \zeta _5}{288}-\frac{2796859 \zeta _7}{4032}+\frac{6767 \pi
              ^6}{5376}-\frac{138163 \pi ^4}{103680}+\frac{181931 \pi
              ^2}{3456}
\nonumber\\&&\mbox{}
-\frac{230063}{768}
\Bigg]
%
%
-\frac{39277 s_{8 a}}{60}-\frac{378593}{486} \pi ^2 \zeta _3^2-\frac{246895
               \zeta _3^2}{2592}+\frac{5465129 \zeta _5 \zeta
               _3}{1080}-\frac{110419 \pi ^4 \zeta _3}{25920}
\nonumber\\&&\mbox{}
-\frac{390271 \pi ^2 \zeta _3}{1296}+\frac{29821 \zeta _3}{18}+\frac{193657
              \pi ^2 \zeta _5}{144}+\frac{12305 \zeta _5}{18}-\frac{7097513
              \zeta _7}{1344}+\frac{64370083 \pi ^8}{163296000}
\nonumber\\&&\mbox{} 
+\frac{2545177 \pi ^6}{4354560}+\frac{586303 \pi ^4}{103680}-\frac{1737749 \pi ^2}{6912}+\frac{7659073}{4608}
%
\nonumber\\&&\mbox{}
+ {\cal O}(\epsilon)\,,
\\
\lefteqn{G_{111111111112}^{\text{(\rm df2-3)}}=}
\nonumber\\&&
+\frac{1}{\epsilon^8}  \Bigg[
-\frac{1}{48}
\Bigg]
%
%
+\frac{1}{\epsilon^7}  \Bigg[
-\frac{5}{8}
\Bigg]
%
%
+\frac{1}{\epsilon^6}  \Bigg[
\frac{5 \pi ^2}{36}-\frac{9101}{1728}
\Bigg]
%
%
+\frac{1}{\epsilon^5}  \Bigg[
\frac{401 \zeta _3}{72}+\frac{104 \pi ^2}{27}-\frac{2683}{648}
\Bigg]
\nonumber\\&&\mbox{}
+\frac{1}{\epsilon^4}  \Bigg[
\frac{21469 \zeta _3}{144}-\frac{19 \pi ^4}{192}+\frac{64867 \pi ^2}{2592}+\frac{1890005}{31104}
\Bigg]
%
%
+\frac{1}{\epsilon^3}  \Bigg[
-\frac{13151}{432} \pi ^2 \zeta _3+\frac{2688043 \zeta _3}{2592}
\nonumber\\&&\mbox{}
+\frac{16277 \zeta _5}{120}-\frac{73759 \pi ^4}{51840}+\frac{52943 \pi ^2}{3888}-\frac{3191177}{23328}
\Bigg]
%
%
+\frac{1}{\epsilon^2}  \Bigg[
-\frac{248513 \zeta _3^2}{432}-\frac{52100 \pi ^2 \zeta _3}{81}
\nonumber\\&&\mbox{}
+\frac{1838789 \zeta _3}{3888}+\frac{103027 \zeta _5}{48}-\frac{751 \pi ^6}{3240}+\frac{82751 \pi ^4}{38880}-\frac{10340263 \pi ^2}{93312}+\frac{7939145}{139968}
\Bigg]
\nonumber\\&&\mbox{}
+\frac{1}{\epsilon}  \Bigg[
-\frac{3037421 \zeta _3^2}{216}+\frac{653 \pi ^4 \zeta _3}{60}-\frac{27028351
              \pi ^2 \zeta _3}{7776}-\frac{328391611 \zeta
              _3}{46656}-\frac{37751 \pi ^2 \zeta _5}{72}
\nonumber\\&&\mbox{}
+\frac{53255227 \zeta _5}{4320}+\frac{2796859 \zeta _7}{1344}-\frac{1650113 \pi ^6}{145152}-\frac{10799 \pi ^4}{1215}+\frac{42718393 \pi ^2}{139968}+\frac{1755738287}{1679616}
\Bigg]
\nonumber\\&&\mbox{}
+ \Bigg[
\frac{39277 s_{8 a}}{20}+\frac{378593}{162} \pi ^2 \zeta _3^2-\frac{327874441
              \zeta _3^2}{3888}-\frac{5465129 \zeta _5 \zeta
              _3}{360}-\frac{3251225 \pi ^4 \zeta _3}{15552}
\nonumber\\&&\mbox{}
-\frac{12190039 \pi ^2 \zeta _3}{5832}+\frac{5299330289 \zeta
              _3}{279936}-\frac{19735721 \pi ^2 \zeta _5}{2160}+\frac{33557879
              \zeta _5}{6480}+\frac{13021045 \zeta _7}{672}
\nonumber\\&&\mbox{}
-\frac{64370083 \pi ^8}{54432000}-\frac{70615283 \pi ^6}{1088640}-\frac{834281549 \pi ^4}{5598720}-\frac{1101059033 \pi ^2}{1679616}-\frac{72028514245}{10077696}
\Bigg]
\nonumber\\&&\mbox{}
+{\cal O}(\epsilon)\,.
\end{eqnarray}
The subscripts denote the exponents of the propagators, where the
order is defined in Fig.~\ref{fig::fam_df2}. The six indices for the
numerators are not shown; they are zero. Furthermore, we have
\begin{eqnarray} 
  s_{8a} &=& \zeta_8 + \zeta_{5,3} \approx 1.0417850291827918834\,.
\end{eqnarray}
$\zeta_{m_{1},\dots,m_{k}}$ are multiple zeta values given by
\begin{eqnarray}
  \zeta_{m_{1},\dots,m_{k}} &=&
  \sum\limits _{i_{1}=1}^{\infty}\sum\limits
  _{i_{2}=1}^{i_{1}-1}\dots\sum\limits _{i_{k}=1}^{i_{k-1}-1}\prod\limits
  _{j=1}^{k}\frac{\mbox{sgn}(m_{j})^{i_{j}}}{i_{j}^{|m_{j}|}}
  \,. 
\end{eqnarray}
Note that $s_{8a}$ cancels in the combination of the master integrals which
leads to the $(d_F^{abcd})^2$ part of the photon quark form
factor, see Eq.~(\ref{eq::Fq_res}).

\end{appendix}



\end{document}